\documentclass[12pt,]{article}
\usepackage{booktabs}
\usepackage{authblk}
\usepackage{fullpage}
\usepackage{amssymb,amsmath}
\usepackage[utf8x]{inputenc}
\usepackage[T1]{fontenc}
\usepackage{siunitx}
\usepackage[version=3]{mhchem}
\usepackage{multirow}

\usepackage[numbers,square,sort&compress]{natbib}
\usepackage{setspace}
\usepackage[unicode=true]{hyperref}
\hypersetup{breaklinks=true,
            bookmarks=true,
            colorlinks=false,
            pdfborder={0 0 0}}
\usepackage{graphicx}
\graphicspath{{figures/}}
\makeatletter
\def\ScaleWidthIfNeeded{%
 \ifdim\Gin@nat@width>\linewidth
    \linewidth
  \else
    \Gin@nat@width
  \fi
}
\def\ScaleHeightIfNeeded{%
  \ifdim\Gin@nat@height>0.9\textheight
    0.9\textheight
  \else
    \Gin@nat@width
  \fi
}
\makeatother
\setkeys{Gin}{width=\ScaleWidthIfNeeded,height=\ScaleHeightIfNeeded,keepaspectratio}%

\title{Data-Driven Forecasting of three-Component Seismograms Using Transformer Architectures}
\author[1]{Waleed Esmail}
\author[2,3]{Stuart Russell}
\author[2]{Jana Klinge}
\author[1]{Alexander Kappes}
\author[2,4]{Christine Thomas}
\affil[1]{Institut f\"ur Kernphysik, Universit\"at M\"unster, Wilhelm-Klemm-Straße 9, 48149, M\"unster}
\affil[2]{Institut f\"ur Geophysik, Universit\"at M\"unster, Corrensstraße 24, 48149, M\"unster}
\affil[3]{James Cook University, 1 James Cook Drive, Douglas, Queensland, 4814, Australia}
\affil[4]{Geological Survey of Denmark and Greenland, Copenhagen, Denmark}

\date{\today}

\begin{document}

\maketitle

\newpage

\section*{Key Points}

\begin{itemize}
    \item \textsc{SeismoGPT} forecasts three-component seismograms with median normalized cross-correlation $\ge$ 0.93 across all evaluation settings.
    \item Forecast performance decreases for large source–receiver distances and small magnitudes, whereas source depth has only a minor influence.
    \item Stable autoregressive forecasting requires observing approximately one S-P interval of post-S-wave motion.
\end{itemize}

\section*{Plain Language Summary}
Future prediction of earthquake ground motion is challenging because seismic waves interact with the Earth's interior in complex ways. In this study, we introduce a machine learning model called \textsc{SeismoGPT} that predicts the later parts of a seismic recording from its earlier portions. To establish a proof of concept, we train and test the model using large datasets of synthetic seismograms generated under controlled conditions. We find that the model can accurately reproduce many features of the future waveforms over a wide range of earthquake sizes and source–receiver distances. Our results show that large machine learning models can learn the underlying patterns of seismic wave propagation and may provide a foundation for future applications in earthquake monitoring, early warning, and other geophysical forecasting problems.

\section*{Abstract}
Forecasting seismic waveforms beyond observed data remains challenging due to the nonlinear, dispersive, and multi-scale nature of seismic wave propagation. In this work, we introduce \textsc{SeismoGPT}, a transformer-based autoregressive model designed to forecast three-component seismic waveforms directly in the time domain.
Forecasting is formulated as a physically constrained continuation problem in which the model receives waveform context beginning at the P-wave arrival and extending a defined time beyond the S-wave arrival, after which future motion is generated recursively without access to ground-truth samples.
Evaluation is performed on synthetic seismograms spanning source depths of 5--100\,km, epicentral distances of 10--90$^\circ$, and magnitudes $3 \leq M_w \leq 7$. To disentangle the effects of context length and prediction horizon, we define three evaluation configurations using a distance-normalized context ratio and fixed prediction horizons of 120 and 240\,s. Across all configurations, the model achieves a median normalized cross correlation of 0.93 or higher.
Analysis of representative forecasts shows that successful predictions preserve both phase coherence and spectral energy distribution. Where failure cases arise, this is primarily due to gradual phase drift during autoregressive rollout rather than unphysical signal generation. These results demonstrate that transformer-based sequence models can learn stable dynamical continuation of seismic wavefields, highlighting the potential of foundation-model approaches for physics-driven time-series forecasting. There are potential applications of this methodology in seismic warning and hazard mitigation, particularly for next-generation gravitational-wave observatories, such as the Einstein Telescope.

\newpage

\section{Introduction}
Accurate and timely forecasting of seismic wavefields is a fundamental component of earthquake monitoring, early-warning, hazard mitigation, and the protection of sensitive infrastructure from seismic waves. In three-dimensional heterogeneous media, seismic waves undergo complex scattering, diffraction, and mode conversions that require numerical treatment \cite{sato2012seismic, igel2017computational}. With current computational resources, high-fidelity simulation of seismic wave propagation remains incredibly computationally demanding, particularly when short periods, long propagation times, large spatial domains or gravitational effects are considered.

Conventional numerical forward modeling approaches, including the Finite-Difference Method \cite{graves1996simulating} and the Spectral Element Method \cite{komatitsch1999introduction}, provide robust deterministic solutions of the elastodynamic equations. However, the computational cost of these methods is dominated by the frequency content of the wavefield and the domain size; the number of mesh elements scales with frequency to the power of 3~-~4, depending on the parameterization used \cite{komatitsch2002spectrala, Nissen2014, leng2016efficient, leng2019axisem3d}, so that while global simulations at long periods are tractable on a single workstation, modeling at periods below $\sim1$~s can exceed the capacity of national-scale computing clusters \cite{xu2024numerical}. This frequency problem is compounded in strongly heterogeneous media, where fine spatial discretization is required for numerical stability and accuracy \cite{graves1996simulating, komatitsch2002spectrala}. Accurate simulation of the longest periods is also computationally problematic due to increasing influence of gravity perturbations, which are treated as negligible at short periods \citep{komatitsch2002spectralb, van2021modelling}. These computational challenges are particularly acute when multiple simulations are required, for example, to characterize the statistical properties of the wavefield or to explore parameter uncertainty. Real-time forecasting at realistic frequencies remains impractical with conventional numerical approaches in most scenarios.

These limitations motivate the exploration of alternative and less demanding computational paradigms for seismogram calculation. In recent years, machine learning approaches have been increasingly investigated as data-driven surrogates for wave propagation and seismic signal analysis \cite{lyu2025rapid}. Machine learning has significantly transformed modern seismology from manual expert inspection into automated analysis of massive global datasets \cite{mousavi2022machine}. Early applications focused primarily on event-level tasks within the earthquake cataloging pipeline \cite{liu2024seislm, kubo2024recent}, including event detection in continuous noisy records \cite{mousavi2020earthquake}, phase picking for precise estimation of P- and S-wave arrival times \cite{mousavi2020earthquake, zhu2019phasenet}, and extraction of seismic signal from anthropogenic or environmental noise \cite{ kong2018machine_myshake, 8802278, heuel2022suppression}.

Methodologically, machine learning in seismology has evolved from shallow multilayer perceptrons to more powerful 1D and 2D convolutional neural networks (CNNs) such as PhaseNet and generalized phase detection (GPD) \cite{zhu2019phasenet, ross2018generalized}, recurrent neural networks (RNNs) including Long Short-Term Memory (LSTMs) and Gated Recurrent Units (GRUs) \cite{liu2024seislm, mojtahedi2025deep}, and hybrid architectures combining CNNs, RNNs, and attention mechanisms, exemplified by the Earthquake Transformer \cite{mousavi2020earthquake}. These models demonstrate strong robustness to noise and often detect substantially more events than traditional automatic pipelines \cite{kong2018machine_insights, kubo2024recent}, achieving picking accuracies comparable to or exceeding those of experienced analysts \cite{kubo2024recent, mousavi2020earthquake}.

More recent developments integrate physical constraints directly into learning frameworks through physics-informed neural networks (PINNs) and related approaches \cite{ren2022seismicnet, kubo2024recent, habib2024applications}, targeting tasks such as 3D elastic wave propagation modeling \cite{ren2022seismicnet}, full waveform inversion, and simulation of earthquake cycles and crustal deformation \cite{habib2024applications}. These methods embed governing partial differential equations into the loss function \cite{kubo2024recent, habib2024applications} or employ neural operators, such as Fourier Neural Operators, to approximate wave equation solvers \cite{kubo2024recent}. Their strengths lie in the incorporation of physical laws and interpretability; yet, training remains computationally demanding and technically challenging for highly nonlinear regimes or sharp discontinuities \cite{habib2024applications}. 

At the forefront, foundation-style models \cite{liu2024seislm, esmail2025forecasting} aim to learn general seismic representations from large-scale unlabeled data through self-supervised objectives such as masked reconstruction \cite{liu2024seislm}, enabling downstream fine-tuning for detection, picking, and classification tasks. Typically implemented as encoder-only transformer architectures adapted from natural language processing models \cite{zhao2023survey}, these systems leverage unlabeled data and perform well in low-data regimes. Nevertheless, seismic foundation modeling remains nascent; scaling behavior is not yet well characterized \cite{liu2024seislm}, and current models primarily operate on single-station waveforms with applications centered on event-level tasks rather than continuous, spatially distributed wavefield forecasting.

In this study, we investigate whether transformer-based sequence models can learn to predict the short-term evolution of seismic waveforms directly from data, without explicit numerical integration of the governing elastodynamic equations. We introduce \textsc{SeismoGPT}, a generative autoregressive transformer-based architecture designed for seismic waveform forecasting. The model captures long-range temporal dependencies through self-attention mechanisms. In contrast to models designed primarily for event classification or representation learning, \textsc{SeismoGPT} is trained explicitly for autoregressive waveform forecasting, enabling direct evaluation of predictive performance over future time windows.

This forecasting framing has direct practical relevance. The model receives context beginning at the P-wave arrival and predicts the subsequent motion, including the surface wave train that carries the majority of seismic energy and causes the greatest structural damage, which addresses the same predictive window exploited by earthquake early warning systems \cite{allen2009EWS}. Although the framework is general, predictive modeling of spatial and temporal seismic ground motion is of particular relevance for next-generation gravitational-wave observatories, such as the Einstein Telescope (ET) \cite{ET:2025xjr}. The ET will be sensitive to gravitational waves at frequencies where its noise budget is dominated by Newtonian noise \cite{Harms:2019dqi}, local fluctuations in the gravitational field caused by the ambient seismic wavefield. Forecasting the short-term evolution of that wavefield could inform active Newtonian noise mitigation.

\section{Forecasting Problem and Data}

\subsection{Problem Definition}
\label{sec:problem_def}
In this study, seismic waveform forecasting is formulated as a time-series prediction task. Given an observed waveform segment of length $T$, denoted by $\mathbf{x}_{1:T}$, the objective is to predict its short-term future evolution $\mathbf{x}_{T+1:T+\Delta}$ over a forecasting horizon $\Delta$. In this work, the input is a single-station three-component (ZNE) seismogram. However, this formulation generalizes naturally to multi-station settings, where the input becomes a collection of synchronized waveform channels forming a spatiotemporal sequence; an array-based extension of this framework is presented in \cite{esmail2025forecasting} in the context of the Einstein Telescope.

Unlike traditional numerical simulation, this approach seeks to learn an implicit data-driven evolution operator that maps past waveform history to its immediate future. The focus of this work is on short-term autoregressive forecasting, where predictions are generated sequentially and fed back into the model for subsequent time steps.

Direct application of transformer architectures to raw seismic time series poses practical challenges. Seismograms are often recorded at high sampling rates of typically 20-200~Hz depending on the intended usage, and may span long durations of minutes or hours, resulting in sequences that are prohibitively long for a standard self-attention mechanism \cite{vaswani2017attention}. Since the computational complexity of attention scales quadratically with sequence length, processing entire continuous records becomes computationally inefficient and memory-intensive.

\begin{figure}[t]
    \centering
    \includegraphics[width=\linewidth]{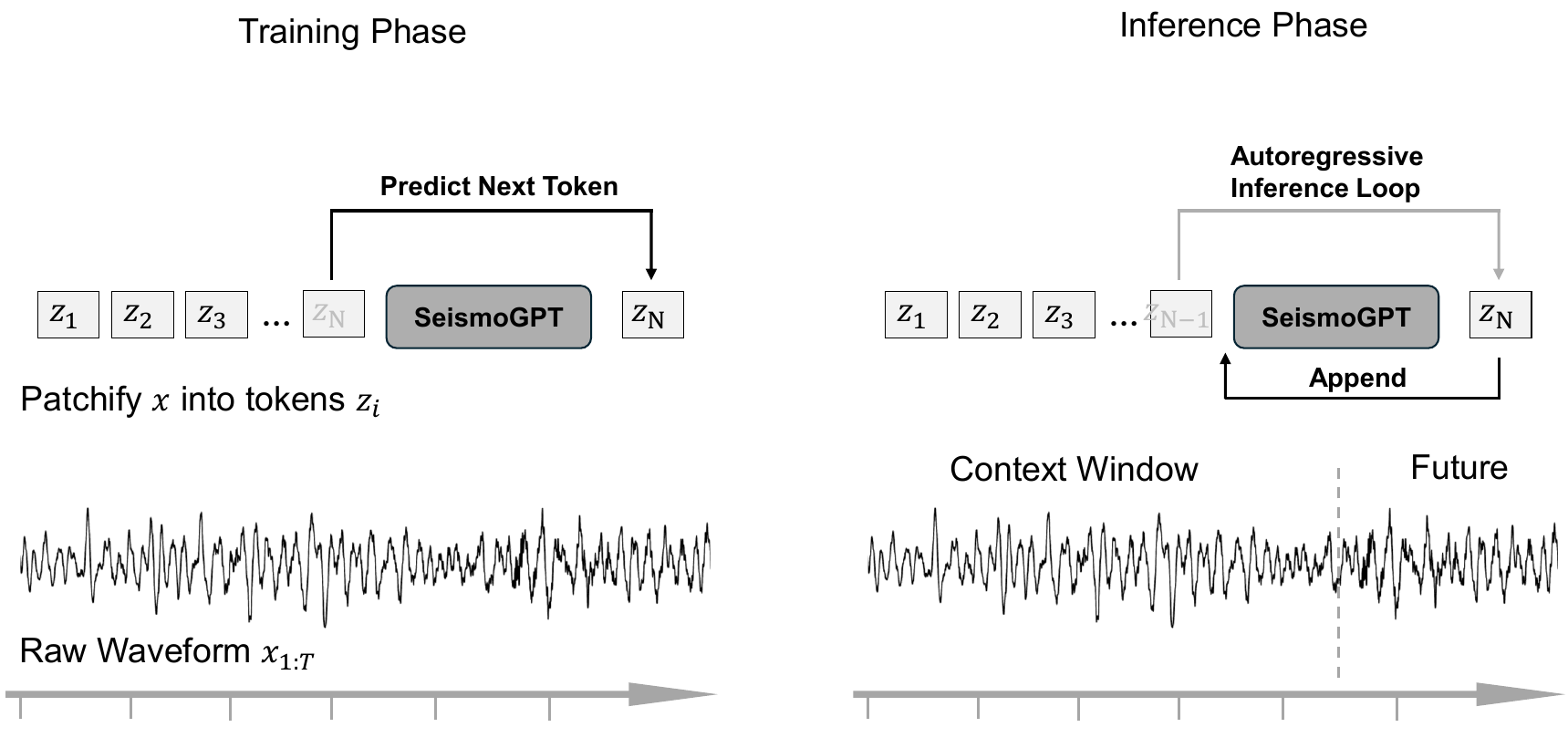}
    \caption{Token-level autoregressive forecasting framework. 
    (left) A continuous seismic waveform segment $\mathbf{x}_{1:T}$ is partitioned into fixed-length tokens $\mathbf{z}_i$ of length $L$ samples through patchification. 
    (right) An autoregressive setup predicts the next token conditioned on previously observed tokens, and predicted tokens are iteratively fed back to generate future waveform segments.}
    \label{fig:autoregressive}
\end{figure}

To address this limitation, the waveform is partitioned into smaller contiguous sub-sequences, or \emph{tokens}, through a patchification process. Each token represents a short temporal segment of the waveform and serves as a fundamental unit for attention-based modeling. This tokenization reduces effective sequence length while preserving local temporal structure, enabling the transformer to capture long-range dependencies across waveform segments rather than individual samples. Similar segmentation strategies have been successfully employed in other domains to scale attention-based models to long signals \cite{talukder2024totem, ansari2024chronos, masserano2024enhancing, persak2024multipleresolution, yang2025tokon}.
\\
\\
In this framework, forecasting operates at the token level rather than at the level of individual samples. Let $\mathbf{z}_1, \mathbf{z}_2, \dots, \mathbf{z}_N$ denote the sequence of waveform tokens obtained by partitioning the continuous signal into fixed-length segments. The forecasting task is formulated autoregressively: given the previously observed tokens, the model predicts the next token in the sequence. Formally, this can be expressed as:

\begin{equation*}
\hat{\mathbf{z}}_{i} = f_{\theta}\!\left(\mathbf{z}_{1}, \mathbf{z}_{2}, \dots, \mathbf{z}_{i-1}\right),
\end{equation*}
where $f_{\theta}$ denotes the transformer parameterized by $\theta$. In this formulation, the model learns an implicit data-driven evolution operator that maps past waveform segments to their short-term future continuation. By operating on tokens rather than individual samples, the sequence length is substantially reduced, improving computational tractability while preserving the ability to model extended temporal dependencies across waveform segments. The resulting token-level autoregressive formulation is illustrated schematically in Figure~\ref{fig:autoregressive}.

To evaluate the feasibility of this formulation in a controlled and reproducible setting, this study only considers synthetic seismic waveforms generated under controlled conditions. This design enables a systematic assessment of predictive stability and error accumulation without the complexities of real seismic data, namely the confounding influence of poorly-constrained source mechanisms, unknown structural heterogeneity, and environmental or anthropogenic noise. Establishing forecasting performance under controlled conditions provides a necessary baseline before the approach can be extended to real-world seismic datasets, potentially through transfer learning \cite{Segura2025seismicprecursors} or domain adaptation strategies \cite{10.1007/978-3-030-71704-9_65}.

\subsection{Synthetic Waveform Generation}
\label{sec:synthetic_generation}

Following the controlled formulation in Section~\ref{sec:problem_def}, we construct a synthetic dataset in which the forward physics, Earth model, source properties, and source-receiver geometry are fully specified. The goal is not to replicate the full realistic complexity of Earth structure and observational noise, but to establish a clean proof-of-concept implementation in a simple semi-realistic case: can an autoregressive transformer learn a stable short-term evolution operator for seismic waveforms under controlled, physically consistent variability? Synthetic seismograms provide three key advantages over real seismic data for this purpose: (i) the generating parameters (Earth model, source mechanism, source depth, and receiver geometry) are known exactly, enabling unambiguous evaluation; (ii) the waveforms can be generated without uncontrolled environmental and instrumental noise, so that forecasting errors can be attributed to the model rather than unknown nuisance factors; and (iii) data volume can be increased systematically to probe scaling with dataset size and waveform diversity, without the sampling biases that would be present when using real recorded seismograms.

A synthetic seismogram can be written schematically as a linear mapping from source-time history and moment tensor components to displacement at a receiver,
\begin{equation}
\mathbf{u}(t;\,\mathbf{x}_r,\mathbf{x}_s) \;=\; \sum_{k=1}^{6} \left(G_k(\mathbf{x}_r,\mathbf{x}_s,t) * s(t)\right) \, m_k,
\end{equation}
where $G_k$ are Green's functions for the chosen Earth model and source-receiver geometry, $m_k$ are the six independent moment-tensor components, $s(t)$ is a source-time function, and $*$ denotes temporal convolution. We generate $\mathbf{u}(t)$ using \texttt{Instaseis} \citep{vanDriel2015}, which synthesizes seismograms from pre-computed Green's function databases computed with \texttt{AxiSEM} \citep{Nissen2014}. \texttt{Instaseis} enables efficient generation of many source-receiver configurations without re-running costly full wavefield simulations for each event. We use the \texttt{ak135f\_2s} database as hosted by Syngine \citep{Hutko2017IRIS, krischer2017syngine}. This database uses the ak135f model \cite{MontagnerKennett1996} and has a minimum period $T_{\min} = 2$ and a sampling rate of $f_s = 1.9$ Hz.

To promote physically realistic waveform diversity, each synthetic event is assigned a centroid moment tensor (CMT) source. Moment tensors are drawn from a distribution fitted to empirical statistics inferred from the Global CMT catalogue \footnote{https://www.globalcmt.org/} \citep{Dziewonski1981b, Ekstrom2012} over a large number of solutions. This yields a family of seismograms that exhibit realistic radiation patterns and magnitude scaling. Each synthetic event is paired with a single receiver at a randomized azimuth and epicentral distance within prescribed ranges (see Table~\ref{tab:synthetic_params}). This one-receiver-per-event design ensures that the event identity and the trace identity coincide, so that a random split on traces produces a clean separation of physical sources across training, validation, and test sets.

\begin{figure}[t]
    \centering
    \includegraphics[width=0.65\linewidth]{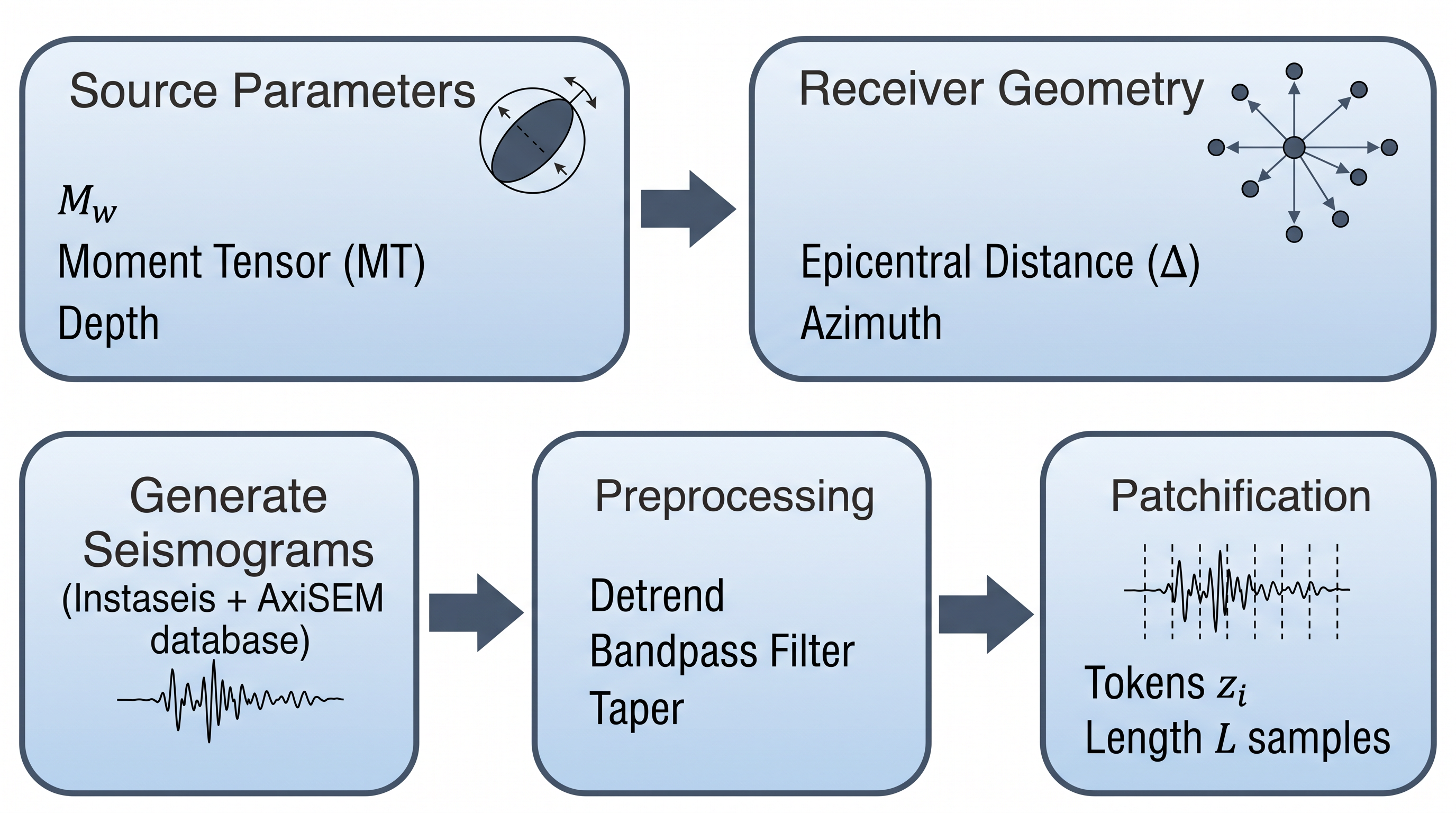}
    \caption{Schematic overview of the synthetic waveform generation pipeline.}
    \label{fig:synthetic_pipeline}
\end{figure}

All synthetic traces are processed with a consistent pipeline to remove trivial trends and isolate the frequency band of interest. Specifically, we apply de-meaning and linear detrending, a taper to suppress edge artifacts, and a fourth-order zero-phase bandpass filter consistent between 2~s, the minimum period of the database, and 100~s, the maximum valid period for AxiSEM Green's function databases due to the treatment of gravity. We also compute the theoretical arrival times of the first-arriving P and S waves using the TauP toolkit \citep{crotwell1999taup} via ObsPy \citep{beyreuther2010obspy}, using the same 1D velocity model as the Green's function database. Finally, each trace is normalized by dividing each component by its peak absolute
amplitude. This per-channel peak normalization removes absolute amplitude differences between events of different magnitudes while preserving the relative amplitude ratios across the three components within each trace. The overall synthetic waveform generation and preprocessing pipeline is summarized schematically in Figure~\ref{fig:synthetic_pipeline}.

Table~\ref{tab:synthetic_params} summarizes the parameters used for synthetic waveform generation and processing. These settings define the controlled variability explored in this work. The complete dataset comprises approximately 3,900,000 three-component seismograms. We partition the data into training, validation, and test sets using an 85/10/5 split, ensuring that no two traces in different splits originate from the same physical source.

\begin{table}[t]
\centering
\caption{Parameters for synthetic waveform generation and preprocessing.}
\label{tab:synthetic_params}
\begin{tabular}{p{0.40\linewidth} p{0.52\linewidth}}
\hline
\textbf{Category} & \textbf{Setting} \\
\hline
\hline
Waveform calculation & \texttt{Instaseis} \citep{vanDriel2015} with Green's function database from \texttt{AxiSEM} \citep{Nissen2014} \\
\hline
Recorded quantity / components & Three-component displacement; (ZNE) \\
\hline
Source type & Full moment tensor (6 components) with CMT-like statistics \citep{Dziewonski1981b, Ekstrom2012} \\
\hline
Magnitude range & $M_w \in [3, 7]$ \\
\hline
Source depth range & $d \in [5, 100]$ km \\
\hline
Epicentral distance range & $\Delta \in [10^\circ, 90^\circ]$ \\
\hline
Total number of traces &  3,900,000  \\
\hline
Train/Val/Test split & [85\%, 10\%, 5\%] \\
\hline
\hline
\end{tabular}
\end{table}

%

\section{\textsc{SeismoGPT} Architecture}
\label{sec:architecture}

Generative Pre-trained Transformer (GPT) models are a class of autoregressive sequence models originally developed for natural language processing \citep{Radford2018ImprovingLU, Radford2019LanguageMA}. Their core principle is simple: given a sequence of discrete tokens\footnote{In natural language processing, a \emph{token} is the fundamental unit of input to the model; typically a word, subword, or character.}, the model is trained to predict the next token by attending only to past context through a causal attention mask. Despite this simplicity, GPT-style architectures have proven remarkably effective at learning complex sequential structures across domains well beyond text, including audio synthesis \citep{borsos2023audiolm}, time-series forecasting \citep{ansari2024chronos}, and protein sequence modeling \citep{ferruz2022protgpt2}.

\begin{figure}[t]
    \centering
    \includegraphics[width=\linewidth]{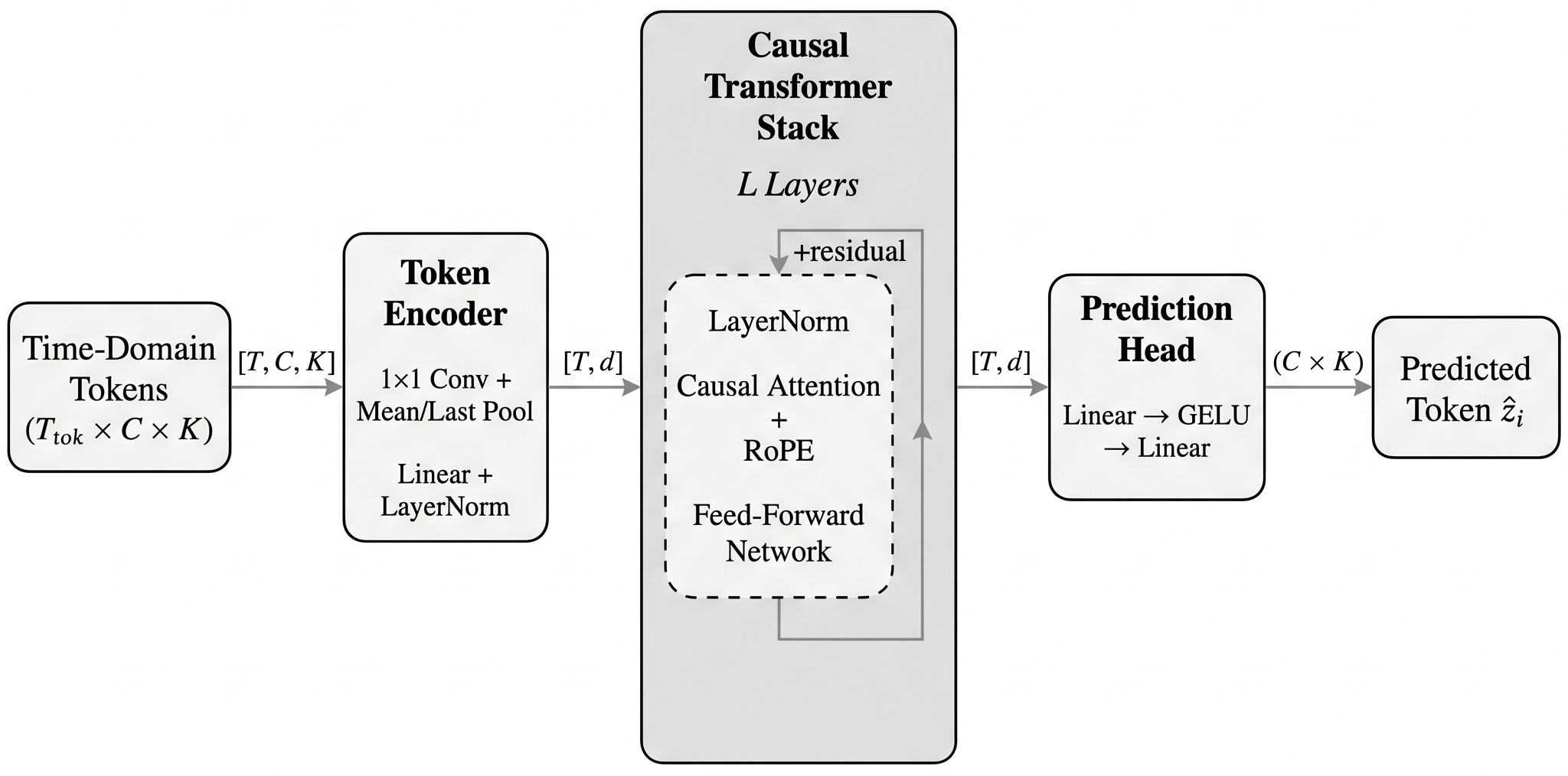}
    \caption{Overview of the \textsc{SeismoGPT} architecture. Input tokens of shape $(T \times C \times K)$ are embedded by the token encoder, which applies a $1{\times}1$ convolution for channel mixing, mean and last-sample pooling over the within-token axis $K$, and a linear projection with layer normalization to produce a sequence of $d$-dimensional token embeddings. These are passed through a stack of $L$ causally masked transformer encoder layers. The prediction head maps each contextual token representation back to waveform space via a two-layer feed-forward neural network, yielding the predicted token $\hat{\mathbf{z}}_i \in \mathbb{R}^{C \times K}$.}
    \label{fig:seismogpt_arch}
\end{figure}

\textsc{SeismoGPT} adapts this paradigm to three-component seismic waveforms. Rather than predicting discrete language tokens, the model operates on continuous-valued waveform patches. The architecture is organized in three stages: a convolutional token encoder that maps waveform patches to fixed-dimensional embeddings, a causal transformer backbone that models temporal dependencies across the token sequence, and a prediction head that maps the learned representations back to waveform space. The overall architecture is summarized in Figure~\ref{fig:seismogpt_arch}.

The remainder of this section describes the token embedding (Section~\ref{subsec:token_embedding}), the transformer backbone (Section~\ref{subsec:transformer_backbone}), the prediction head (Section~\ref{subsec:prediction_head}), and the training strategy (Section~\ref{sec:training}).

\subsection{Token Embedding}
\label{subsec:token_embedding}
As described in Section~\ref{sec:problem_def}, the three-component seismogram is partitioned into a sequence of fixed-length tokens $\mathbf{z}_i \in \mathbb{R}^{C \times K}$, each containing $K$ consecutive samples across all $C$ components. Before these tokens can be processed by the transformer backbone, each one must be mapped to a fixed-dimensional vector $\mathbf{e}_i \in \mathbb{R}^{d}$. This is done through two operations. First, a pointwise ($1 \times 1$) convolution mixes the three seismogram components at each sample position independently, projecting from $C$ channels to $d$ features. This step allows the model to learn component interactions; for instance, the relative amplitude and phase relationships between vertical and horizontal channels. Second, the resulting $d \times K$ feature map is summarized into a single $d$-dimensional vector by concatenating the mean over the $K$ sample positions with the feature vector at the last sample position, followed by a linear projection and layer normalization. The concatenation of mean and last-sample features provides the embedding with both an average representation of the token and a snapshot of its most recent state, which is particularly relevant for the causal forecasting task where the boundary between tokens carries forward in time.

This embedding operates independently on each token; that is, no information is exchanged between tokens at this stage. Cross-token temporal dependencies are modeled entirely by the causal transformer backbone described in the next section. The full sequence of $N$ independent tokens $\mathbf{E} = [\mathbf{e}_1, \dots, \mathbf{e}_N] \in \mathbb{R}^{N \times d}$ is then passed to the transformer encoder.

\subsection{Causal Transformer Backbone}
\label{subsec:transformer_backbone}

The token embedding sequence $\mathbf{E} = [\mathbf{e}_1, \dots, \mathbf{e}_N] \in \mathbb{R}^{N \times d}$ is processed by a stack of $L$
identical transformer encoder layers with causal self-attention \citep{vaswani2017attention}. We adopt an encoder-only architecture rather than a full encoder-decoder configuration, because the forecasting task is single-sequence autoregressive modeling: both input and output lie in the same token space, with future waveform tokens conditioned solely on
past tokens. There is no separate conditioning stream that would require cross-attention, making a causally masked encoder stack both sufficient and computationally efficient.

Each layer applies, in sequence, layer normalization \citep{Baetal2016}, multi-head self-attention with a causal mask that prevents each token from attending to future tokens, and a position-wise feed-forward network, with residual connections around both the attention and feed-forward blocks (pre-norm configuration).

To encode temporal ordering, we apply Rotary Positional Embeddings (RoPE) \citep{su2024roformer} to the query and key attention vectors before attention score computation. RoPE encodes relative position through phase rotations in feature space, which allows the model to generalize to sequence lengths beyond those seen during training without introducing additional learned parameters. Stacking $L$ such layers yields hierarchical modeling of long-range temporal dependencies across waveform tokens.

\subsection{Prediction Head}
\label{subsec:prediction_head}

The final token representations $\mathbf{H}_L \in \mathbb{R}^{N \times d}$ are projected back to waveform space by a prediction head consisting of two linear layers with a Gaussian Error Linear Unit (GELU) activation \cite{HendrycksGimpel2016}:

\begin{equation}
\hat{\mathbf{z}}_i = \mathbf{W}_2 \,
\mathrm{GELU}\!\left(\mathbf{W}_1 \, \mathbf{h}_i^{(L)}
+ \mathbf{b}_1\right) + \mathbf{b}_2
\;\in\; \mathbb{R}^{C \times K},
\end{equation}

\noindent where $\mathbf{h}_i^{(L)} \in \mathbb{R}^{d}$ is the output of the final transformer layer for token $i$. This maps from the abstract token embedding space back to the physical three-component waveform domain, producing the next-token prediction $\hat{\mathbf{z}}_i \in \mathbb{R}^{C \times K}$.

\subsection{Training Strategy}
\label{sec:training}
The model is trained in a supervised setting using teacher forcing: at each step, the ground-truth past tokens are provided as context, and the model predicts the next token. Each token spans $K = 16$ samples at the native sampling rate of $\approx$1.9\,Hz, corresponding to approximately 8.4\,s of ground motion. A context window of $N = 320$ tokens would therefore cover roughly 45 minutes of continuous waveform. This section describes the training objective and its individual components.

\subsubsection{Loss Function}
Training seismic waveform models with a purely time-domain loss can produce predictions that match the target sample-by-sample but fail to reproduce its spectral content, for example, smearing dispersive surface-wave trains or shifting dominant frequencies. Conversely, a purely spectral loss does not penalize timing errors in individual phases. To address both aspects, we combine a time-domain loss with a multi-resolution spectral loss and two auxiliary regularization terms.

The primary time-domain objective is the log-cosh loss,

\begin{equation}
\mathcal{L}_{\mathrm{time}} = \frac{1}{N C K} \sum_{i,c,k}
\log \cosh\!\left(\hat{z}_{i,c,k} - z_{i,c,k} \right),
\end{equation}

\noindent where $\hat{z}_{i,c,k}$ and $z_{i,c,k}$ are the predicted and target values at token $i$, component $c$, and sample $k$. The log-cosh function behaves like mean-squared error for small residuals and like mean-absolute error for large ones, combining sensitivity to small mismatches with robustness to the heavy-tailed residuals that are characteristic of seismic waveforms. Unlike L1 loss, it is smooth and differentiable everywhere, which improves optimization stability.

In addition, to encourage spectral fidelity, we add a multi-resolution short-time Fourier transform (STFT) loss \citep{yamamoto2020parallel}. For each of three FFT
window sizes $\{128, 256, 512\}$, we compute the STFT magnitude spectra of the predicted and target waveforms and take their mean absolute difference:

\begin{equation}
\mathcal{L}_{\mathrm{STFT}} = \frac{1}{|\mathcal{F}|}
\sum_{n \in \mathcal{F}}
\left\lVert
\left|\mathrm{STFT}_n(\hat{\mathbf{z}})\right|
- \left|\mathrm{STFT}_n(\mathbf{z})\right|
\right\rVert_1,
\end{equation}

\noindent where $\mathcal{F} = \{128, 256, 512\}$ denotes the set of FFT sizes. A single FFT size imposes a fixed time--frequency trade-off; using multiple resolutions avoids this by capturing spectral structure at different scales. Short windows (128 samples) penalize errors in transient content such as body-wave onsets, while longer windows (512 samples) are more sensitive to the lower-frequency character of surface waves and coda.

A temporal delta loss penalizes differences in the token-to-token transitions between prediction and target:

\begin{equation}
\mathcal{L}_{\delta}
= \frac{1}{(N{-}1) C K}
\sum_{i=2}^{N} \sum_{c,k}
\left(
\Delta\hat{z}_{i,c,k} - \Delta z_{i,c,k}
\right)^2,
\end{equation}

\noindent where $\Delta\hat{z}_{i} = \hat{z}_{i} - \hat{z}_{i-1}$ and likewise for the target. This term encourages smooth transitions across token boundaries, which is particularly important during autoregressive rollout where small boundary discontinuities can accumulate over many steps.

Finally, during training the model predicts not only the immediate next token but $H = 4$ future tokens simultaneously. This is achieved by adding a learnable horizon embedding $\mathbf{e}_h \in \mathbb{R}^{d}$ to the transformer output before passing it through the shared prediction head, giving each horizon its own specialization without duplicating parameters. The loss is computed independently for each horizon and summed with geometrically decaying weights $w_h = \gamma^{h-1}$ (with $\gamma = 0.6$, giving weights $1.0, 0.6, 0.36, 0.22$ for horizons 1 through 4), so that the primary next-token prediction dominates the gradient while the
auxiliary horizons encourage the encoder to represent longer-range temporal structure. At inference, only the first-horizon prediction ($h = 1$) is used. To enforce spectral consistency across adjacent prediction horizons, a cross-horizon coherence loss applies the multi-resolution STFT loss to concatenated pairs of consecutive horizon predictions versus concatenated ground truth.

The total training objective is

\begin{equation}
\mathcal{L}
= \sum_{h=1}^{H} w_h
\left(
\mathcal{L}_{\mathrm{time}}^{(h)}
+ \lambda_{\mathrm{STFT}}\,
\mathcal{L}_{\mathrm{STFT}}^{(h)}
+ \lambda_{\delta}\,
\mathcal{L}_{\delta}^{(h)}
\right)
+ \lambda_{\mathrm{coh}}\,
\mathcal{L}_{\mathrm{coh}},
\end{equation}

\noindent with the weighting variables $\lambda_{\mathrm{STFT}}$, $\lambda_{\delta}$, and $\lambda_{\mathrm{coh}}$ being hyperparameters. For the rest of this paper, we use $\lambda_{\mathrm{STFT}} = 0.05$, $\lambda_{\delta} = 0.05$, and $\lambda_{\mathrm{coh}} = 0.1$.

A detailed ablation study and methodological analysis of the horizon-embedding prediction head and cross-horizon coherence loss are presented in \cite{esmail_methods_paper_2026}.

\subsubsection{Optimization}
All experiments are conducted using PyTorch \citep{paszke2019pytorch} and PyTorch Lightning \citep{falcon2019pytorch} on the PALMA~II cluster at the University of M\"unster\footnote{\url{https://palma.uni-muenster.de/}} using NVIDIA RTX~4090 GPUs with bf16 mixed-precision training. The model is optimized with AdamW \citep{loshchilov2017decoupled} at an initial learning rate of $10^{-4}$, with a linear warmup over 1000 steps followed by cosine annealing with warm restarts. Training runs for up to 50 epochs with early stopping (patience 10) monitoring the validation loss. All model and training hyperparameters are summarized in Table~\ref{tab:training_config}.

\begin{table}[t]
\centering
\caption{Model architecture and training configuration.}
\label{tab:training_config}
\begin{tabular}{p{0.45\linewidth} p{0.45\linewidth}}
\hline
\textbf{Parameter} & \textbf{Value} \\
\hline
Total parameters & $\approx$ 26 M \\
Embedding dimension $d$ & 512 \\
Transformer layers $L$ & 8 \\
Attention heads & 8 \\
Feed-forward multiplier & 4 \\
Positional encoding & RoPE \\
Dropout & 0.1 \\
Token size $K$ & 16 samples ($\approx$ 8.4 s) \\
Training Context length $N$ & 320 tokens ($\approx$ 45 min) \\
Prediction horizons $H$ & 4 \\
Horizon weight decay $\gamma$ & 0.6 \\
$\lambda_{\mathrm{STFT}}$ / $\lambda_{\delta}$ / $\lambda_{\mathrm{coh}}$ & 0.05 / 0.05 / 0.1 \\
Optimizer & AdamW \\
Learning rate & $1 \times 10^{-4}$ \\
LR schedule & Warmup + cosine annealing \\
Max epochs & 50 \\
Early stopping & Patience 10, validation loss \\
\hline
\end{tabular}
\end{table}

\section{Results}

\subsection{Physical Framing and Evaluation Protocol}

We evaluate the forecasting capability of the \textsc{SeismoGPT} architecture in a physically constrained setting defined relative to the seismic phase arrivals. Forecasting performance depends on two factors: how much of the waveform the model has observed and how far ahead it must predict. To disentangle these, we define configurations with different context windows, defined relative to seismic phase arrivals, and prediction horizons, that are fixed in seconds. In this study, we focus exclusively on forecasting the post-S-wave seismogram, with both the P- and S-wave given as context. 

For each seismogram, the context window begins at the ray-theoretical P-wave arrival time and extends beyond the ray-theoretical S-wave arrival time by an offset equal to $r$ times the S-P travel-time difference:

\begin{equation}
t_{\mathrm{ctx}} = r \times (t_S - t_P),
\end{equation}

\noindent where $t_P$ and $t_S$ are the ray-theoretical P- and S-wave arrival times and $r$ is the context ratio. As the P- and S-wave separation depends on epicentral distance and source depth, this design ensures that the model always observes the same physically meaningful arrival-referenced wavefield evolution regardless of source-receiver geometry, without resorting to arbitrarily defined time windows. At $r = 1$, the context extends from the P-wave arrival to the S-wave arrival, so the model observes the P wave, the P-coda, and the S-wave onset. At $r = 2$, the context extends one additional S--P interval beyond the S-wave arrival, encompassing the early surface-wave train and the beginning of the coda.

The prediction horizon, by contrast, is specified as a fixed time duration $\Delta t_{\mathrm{fut}}$ in seconds after the end of the context window, so that the forecasting difficulty can be compared directly across events at different distances. From the end of the context window to the prediction horizon, the model generates the future waveform in fully autoregressive mode, which means that each predicted token is fed back as input for the next step, with no access to future ground truth.

We evaluate three configurations, summarized in Table~\ref{tab:forecast_configs}, designed to isolate the effects of prediction horizon and context length. Comparing A and B (same context, doubled horizon) reveals how performance degrades with forecast length. Comparing B and C (same horizon, doubled context) reveals whether additional post-S observations improve long-horizon forecasts. All results are reported on the hold-out test set spanning source depths of 5--100 km, epicentral distances of 10--90${^\circ}$, and magnitudes $3 \leq M_w \leq 7$. We define three different performance measures to assess different key seismogram characteristics: time-domain timing and phase, time-domain amplitudes, and spectral content.

\begin{table}[t]
\centering
\caption{Forecasting configurations. The context window begins at the P-wave arrival and extends to $r \times (t_S - t_P)$; the model then predicts $\Delta t_{\mathrm{fut}}$ seconds of waveform in autoregressive rollout.}
\label{tab:forecast_configs}
\renewcommand{\arraystretch}{1.3}
\begin{tabular}{@{} c c c @{}}
\hline
Configuration & Context ratio $r$ &
Prediction horizon $\Delta t_{\mathrm{fut}}$ \\
\hline
A & 1.0 & 120\,s \\
B & 1.0 & 240\,s \\
C & 2.0 & 240\,s \\
\hline
\end{tabular}
\end{table}

Time-domain waveform similarity between the predicted and true future segments is quantified using the normalized cross-correlation (NCC) \citep{papasotiriou2022similarity, bobrov2014perspectives}, defined as

\begin{equation}
\mathrm{NCC}(\mathbf{y},\hat{\mathbf{y}})
= \frac{\langle \mathbf{y},\,\hat{\mathbf{y}} \rangle}{\|\mathbf{y}\|_2 \;\|\hat{\mathbf{y}}\|_2 + \epsilon},
\end{equation}

\noindent where $\mathbf{y}$ and $\hat{\mathbf{y}}$ denote the true and predicted waveform segments over the prediction horizon and $\epsilon$ is a small constant for numerical stability. NCC, which is bounded between $-1$ and $1$, is scale-invariant, and therefore emphasizes waveform timing and shape and phase agreement rather than absolute amplitude.

To complement the NCC, we also report the signal-to-residual ratio (SRR), which captures amplitude fidelity by treating the prediction residual as noise:

\begin{equation}
\mathrm{SRR}(\mathbf{y}, \hat{\mathbf{y}}) = 10 \, \log_{10}
\frac{\frac{1}{N}\sum_{i} y_i^2}{\frac{1}{N}\sum_{i} (\hat{y}_i - y_i)^2 + \epsilon},
\end{equation}

\noindent where the sums run over all samples in the prediction horizon. Higher values indicate better predictions; an SRR of 10\,dB means the signal power exceeds the residual power by a factor of ten. Unlike NCC, SRR is sensitive to both amplitude and phase errors, providing a complementary view of forecast quality.

In the final measure, spectral fidelity is assessed by comparing the Welch
power spectral densities (PSD) \citep{welch1967use} of the predicted and true segments over the prediction horizon. To isolate the relative frequency distribution from absolute power level, each PSD is normalized to unit total area prior to comparison. We report the log-spectral $L^2$ error,

\begin{equation}
\mathcal{E}_{\mathrm{PSD}} = \frac{1}{N_f} \sum_{k=1}^{N_f} \left(\log_{10} \tilde{P}_{\hat{y}}(f_k) - \log_{10} \tilde{P}_{y}(f_k) \right)^2, \qquad \tilde{P}(f_k) = \frac{P(f_k)}{\sum_{j} P(f_j) + \epsilon},
\end{equation}

\noindent where $P_{y}$ and $P_{\hat{y}}$ are the power spectral densities of the true and predicted waveforms and $\tilde{P}$ denotes the area-normalized density. This normalization ensures that the metric measures how well the model reproduces the distribution of energy across frequencies, independent of the overall signal amplitude. Lower values indicate better spectral agreement.

\subsection{Overall Forecasting Performance}
\label{sec:overall_performance} 
\begin{figure*}[t]
\centering
\includegraphics[width=\textwidth]{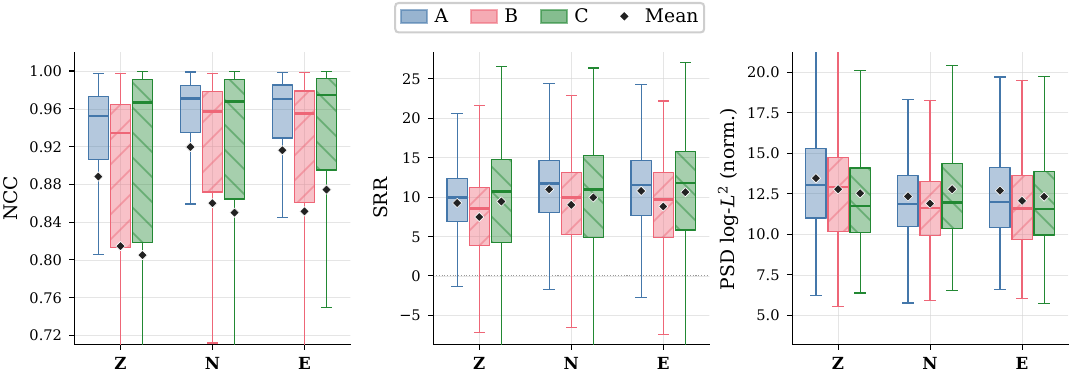}
\caption{Distribution of NCC (left), SRR (center), and normalized PSD log-$L^2$ error (right) on the prediction horizon for the three waveform components (ZNE) across configurations A--C. Box plots show the median, interquartile range, and outliers; diamond markers indicate the mean.}
\label{fig:metrics_boxplot}
\end{figure*}

We first assess the overall forecasting quality across all three configurations using the NCC, SRR, and PSD log-$L^2$ error computed over the prediction horizon. Figure~\ref{fig:metrics_boxplot} summarizes the per-component distributions for the full test set, and Table~\ref{tab:summary_metrics} reports the corresponding median and mean values.

\begin{table}[t]
\centering
\caption{Summary of forecasting metrics across configurations. Values are reported as median (mean) $\pm$ standard deviation over the test events.}
\label{tab:summary_metrics}
\renewcommand{\arraystretch}{1.25}
\begin{tabular}{@{} cl ccc @{}}
\hline
Metric & Config & Z & N & E \\
\hline
\multirow{3}{*}{NCC}
  & A & 0.95 (0.89) $\pm$ 0.19 & 0.97 (0.92) $\pm$ 0.16 & 0.97 (0.92) $\pm$ 0.16 \\
  & B & 0.93 (0.81) $\pm$ 0.27 & 0.96 (0.86) $\pm$ 0.24 & 0.96 (0.85) $\pm$ 0.26 \\
  & C & 0.97 (0.81) $\pm$ 0.36 & 0.97 (0.85) $\pm$ 0.28 & 0.97 (0.87) $\pm$ 0.25 \\
\hline
\multirow{3}{*}{SRR (dB)}
  & A & 9.9 (9.2) $\pm$ 4.7   & 11.7 (11.0) $\pm$ 5.1  & 11.5 (10.8) $\pm$ 5.3  \\
  & B & 8.5 (7.5) $\pm$ 5.1   & 10.0 (9.0) $\pm$ 5.4   & 9.7 (8.8) $\pm$ 5.6    \\
  & C & 10.7 (9.4) $\pm$ 6.9  & 11.0 (10.0) $\pm$ 6.8  & 11.8 (10.6) $\pm$ 6.7  \\
\hline
\multirow{3}{*}{\shortstack{PSD log-$L^2$}}
  & A & 13.0 (13.5) $\pm$ 3.4 & 11.9 (12.3) $\pm$ 2.7  & 12.0 (12.7) $\pm$ 3.2  \\
  & B & 12.9 (12.8) $\pm$ 3.5 & 11.6 (11.9) $\pm$ 2.9  & 11.6 (12.1) $\pm$ 3.4  \\
  & C & 11.7 (12.5) $\pm$ 3.4 & 11.9 (12.8) $\pm$ 3.6  & 11.5 (12.3) $\pm$ 3.6  \\
\hline
\end{tabular}
\end{table}

Across all configurations the model achieves high median NCC values, ranging from 0.93 to 0.97 depending on component and scenario. While all components are predicted well, the horizontal components (N and E) are predicted slightly better than the vertical (Z). This likely reflects both a physical and a statistical effect: the vertical component is more sensitive to complex interference between body-wave reverberations and surface-wave mode conversions, and the model has two horizontal channels to learn from in every trace but only one vertical, leading to greater phase variability. The same pattern is reflected in the SRR, where the horizontal components consistently exceed the vertical by 1--2\,dB.

The comparison between configurations isolates the effects of prediction horizon and context length. Going from A to B; doubling the prediction horizon from 120\,s to 240\,s while keeping the context fixed at $1 \times (t_S - t_P)$ reduces the median NCC by approximately 2 percentage points and the median SRR by 1.5--2\,dB across all components. This degradation is expected as longer autoregressive rollouts accumulate prediction errors, and the later portions of the coda carry less coherent energy, making them intrinsically harder to forecast.

Going from B to C; doubling the context from $1 \times (t_S - t_P)$ to $2 \times (t_S - t_P)$ while keeping the same 240\,s prediction horizon recovers much of this lost performance. The median NCC returns to values comparable to or exceeding Configuration~A (e.g., 0.97 for Z in C versus 0.95 in A), and the median SRR improves by approximately 2\,dB relative to B. This demonstrates that additional post-S context provides the model with further useful information about the evolving wavefield structure, enabling more stable long-horizon forecasts.

A notable feature across all configurations is the gap between the median and mean NCC. In Configuration~A, the mean NCC for the Z component is 0.89 versus a median of 0.95; in Configuration~C, this gap widens to 0.81 versus 0.97. This indicates a tail of poorly predicted events that pulls the mean down while leaving the median largely unaffected. The increasing standard deviation from A to C (Table~\ref{tab:summary_metrics}) confirms that the longer context, while beneficial for typical events, does not prevent occasional autoregressive divergence. Identifying the physical characteristics of these failure cases is the subject of Section~\ref{sec:results_geometry}.

The PSD log-$L^2$ error, which isolates spectral shape from the absolute power level, shows a pattern consistent with the other two metrics. All three configurations produce similar spectral shape errors (medians in the range 11.5--13.0), with Configuration~C achieving the lowest values (median 11.5--11.9 across components) and Configurations~A and B being comparable (medians around 11.6--13.0). This confirms that additional post-S context improves not only phase and amplitude agreement but also the fidelity of the predicted frequency distribution. The consistency across all three metrics, NCC, SRR, and PSD, strengthens the conclusion that Configuration~C produces the best overall forecasts.

\subsection{Dependence on Propagation Geometry}
\label{sec:results_geometry}

\begin{figure*}
\centering
\includegraphics[width=\textwidth]{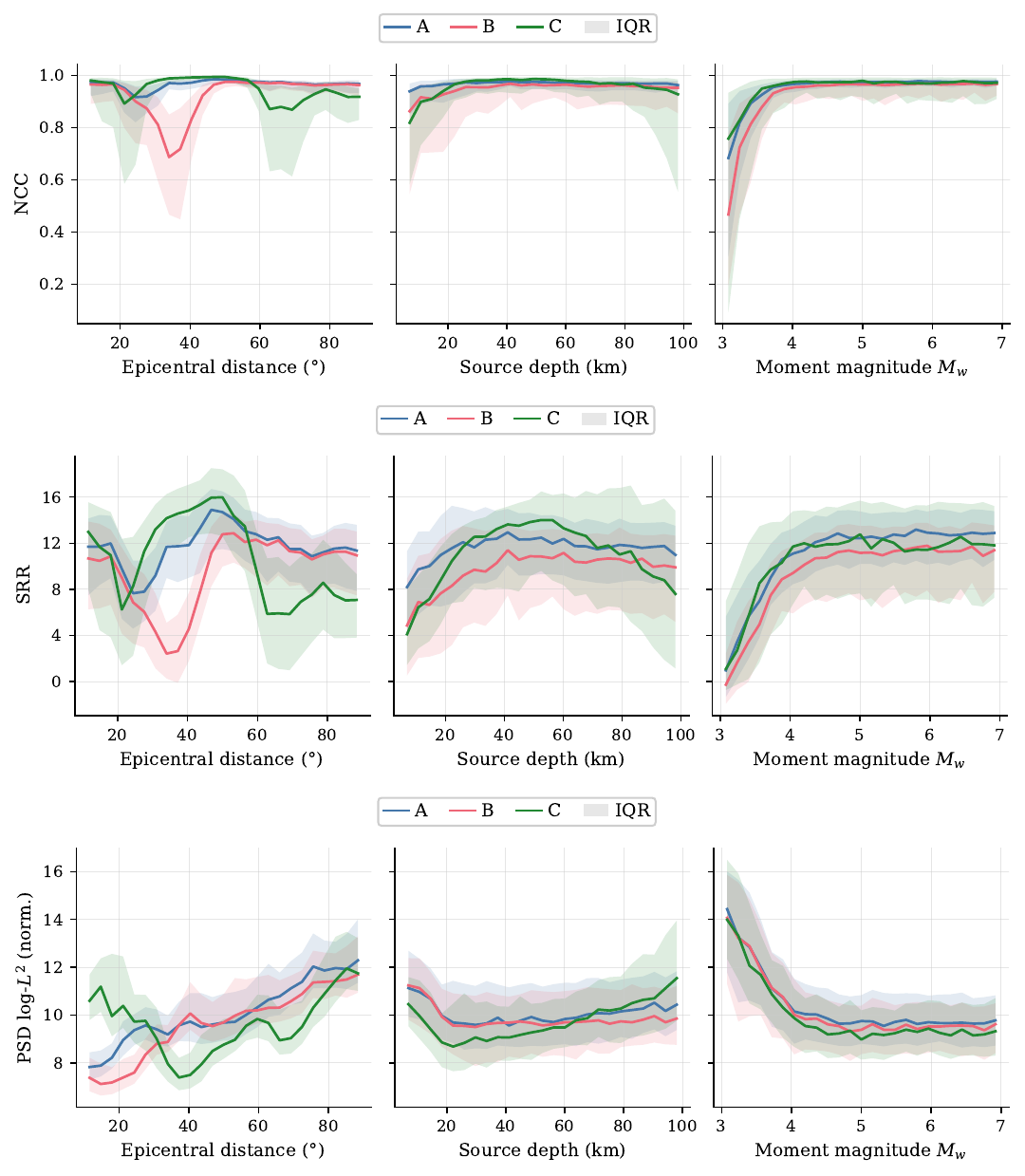}
\caption{Dependence of NCC (top row), SRR (middle row) and normalized PSD log-$L^2$ error (bottom row) on epicentral distance (left column), source depth (center column), and moment magnitude (right column). Solid lines show the running median for configurations A--C and shaded bands indicate the interquartile range.}
\label{fig:ncc_geometry}
\end{figure*}

The aggregate metrics in the previous section average over events with very different physical characteristics. To understand where the model performs well and where it struggles, we examine how NCC, SRR, and PSD log-$L^2$ error depend on the three key parameters that define the source-receiver geometry: epicentral distance, moment magnitude, and source depth. Figure~\ref{fig:ncc_geometry} shows the running median of each metric with interquartile-range shading for all three configurations as a function of each parameter. Furthermore, heatmaps showing how the different performance metrics correlate between the different parameters are shown in Appendix \ref{app:heatmaps}. Each parameter is discussed in turn below.

\subsubsection{Moment magnitude}
Generally, forecast quality improves with increasing magnitude, with the steepest improvement occurring between $M_w \approx 3$ and $M_w \approx 4$. Above $M_w \approx 4$, the median NCC saturates near 0.95 and the IQR narrows considerably. The same overall trend of increasing performance and saturation is displayed by the SRR. The normalized PSD log-$L^2$ error follows a similar trend, with slightly higher spectral shape errors at low magnitudes that decrease and stabilize above $M_w \approx 4$, consistent with the NCC and SRR.

This trend is physically straightforward to interpret: larger earthquakes generate higher-energy waveforms with more coherent phase structure, providing a stronger and more predictable signal for the model to extrapolate. The saturation at high magnitudes suggests that above a certain signal level, performance is limited by the complexity of the wavefield evolution rather than by signal strength.

\subsubsection{Source depth}
The dependence on source depth is also relatively simple. The NCC is typically high across all source depths, but does decrease slightly for the shallowest source depths below approximately 20~km. The trend in SRR is similar but notably more pronounced, indicating that it is specifically amplitude information that is not being well-predicted. The normalized PSD log-$L^2$ error is approximately flat across depths, although again there is an increase in error for the shallowest source depths. That shallow depths are problematic is not surprising, due to the more complex nature of the seismic wavefield at shallow source depths: body wave phases have overlapping depth phases and the surface waves will be comparably much higher amplitude.

\subsubsection{Epicentral distance}
Distance presents the most complex trends of three parameters, with clear differences between the three configurations, requiring a thorough analysis including the correlations between parameters. NCC remains high, but has configuration-specific fluctuations that are accompanied by broadening of the interquartile range. Figure \ref{fig:corner_b} shows that the trough in NCC at 30-40\textdegree\ distance for configuration B, occurs across all source depths and magnitudes, but is much stronger for shallow events with low magnitude. Similarly, the trough in NCC for at 60-70\textdegree\ distance for configuration C is also concentrating in shallow events with low magnitude. These same trends are mirrored by the SRR. 

All of the measures see a slight decrease in performance at longer distances, which is unsurprising as the wavefield becomes increasingly dominated by dispersive surface waves with complex group-velocity structure, and the longer propagation path introduces stronger attenuation and multipathing. These effects make the waveform intrinsically harder to predict, as small errors accumulate over the extended surface-wave train. The normalized PSD log-$L^2$ error mirrors these trends, with increasing spectral shape errors at larger distances for all three configurations.

\subsubsection{Identifying the failure tail}
The mean--median gap noted in the previous section can now be attributed primarily to low-magnitude events at large epicentral distances. These represent the physically most challenging regime: weak signals propagating over long paths, producing low-energy, highly dispersive waveforms that are difficult to forecast autoregressively. The fact that the model's failure cases concentrate in this physically demanding corner of the parameter space, rather than occurring randomly, provides evidence that the learned representation captures meaningful aspects of seismic wave propagation.

\begin{figure*}[!bht]
\centering
\includegraphics[width=\textwidth]{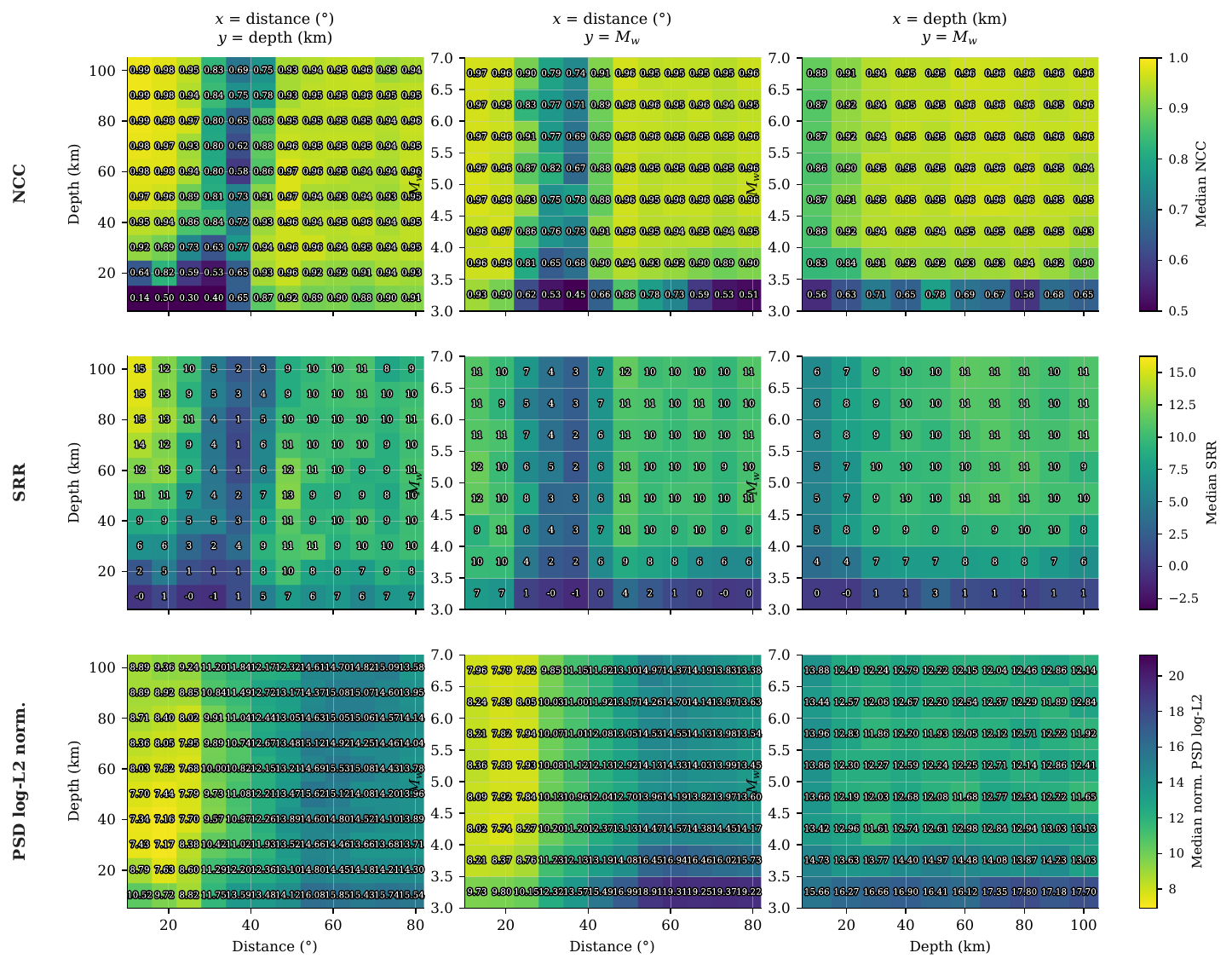}
\caption{Two-dimensional parameter-plane maps of median NCC (top), SRR (middle), and normalized PSD log-$L^2$ error (bottom) for Configuration~B.}
\label{fig:corner_b}
\end{figure*}


To confirm that the degradation reflects a structured failure regime rather than a simple one-dimensional dependence on any single parameter, we computed median metrics over two-dimensional parameter planes. We focus on Configuration~B (Figure~\ref{fig:corner_b}), where trends with the three parameters are most pronounced (Figure~\ref{fig:ncc_geometry}); the corresponding maps for Configurations~A and C are provided in Appendix~\ref{app:heatmaps} and show qualitatively consistent patterns with reduced severity~(A) or a shifted failure region~(C). The maps for Configuration~B reveal that the reduced NCC is not simply a function of increasing distance but is concentrated in specific parameter combinations. The most severe degradation occurs for shallow, low-magnitude events ($M_w \lesssim 4.5$, $d \lesssim 30$\,km), both at short distances ($\Delta \lesssim 20^\circ$) and at the largest distances ($\Delta \gtrsim 70^\circ$). At intermediate and large distances the degradation is driven by dispersive surface-wave complexity, while at short distances it likely reflects the more impulsive, high-frequency character of the near-field wavefield, which is harder to extrapolate with limited post-S context. Events with $M_w \gtrsim 5$ maintain median NCC values above 0.95 across the full distance range.

This pattern is physically interpretable. Low-magnitude, shallow events at large distances produce weak, highly dispersive waveforms in which the post-S wavefield is dominated by surface-wave energy with complex group-velocity structure. In Configuration~B, the model is released after only $1 \times (t_S - t_P)$ of post-S context but must generate 240\,s autoregressively, insufficient to constrain the dispersive phase evolution before the model must predict through it. For stronger or closer events, the signal coherence is high enough to sustain stable prediction even with limited context, while for the most challenging parameter combinations the model lacks both signal energy and observational constraint.

\begin{figure}[t]
\centering
\includegraphics[width=0.75\linewidth]{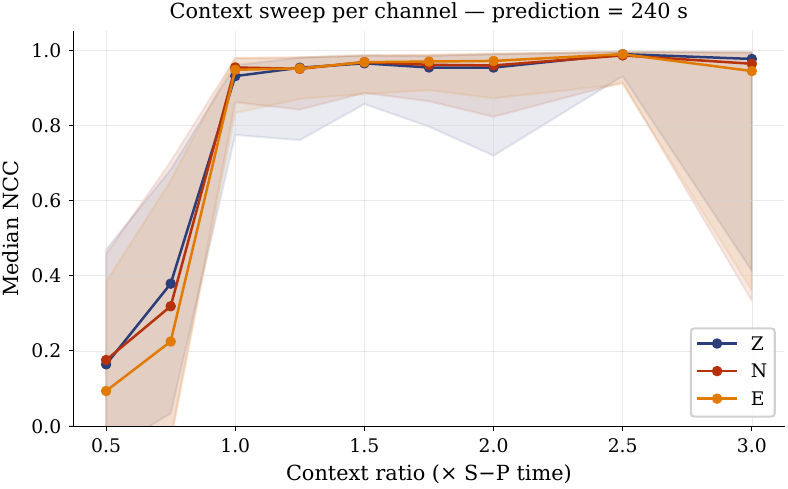}
\caption{Median NCC as a function of context ratio for a fixed prediction horizon of 240\,s, shown per component (Z, N, E). Shaded bands indicate the interquartile range. Performance rises sharply at $r = 1$ and saturates for $r \geq 1.5$, confirming that the S-wave arrival and early surface-wave energy are critical for stable forecasting.}
\label{fig:context_sweep}
\end{figure}

The context-ratio sweep (Figure~\ref{fig:context_sweep}) provides independent confirmation of this interpretation. For context ratios below $1 \times (t_S - t_P)$, the median NCC is low across all components, indicating that the model has not yet observed enough of the post-S wavefield to infer a stable continuation. Performance increases sharply at $r = 1$, coinciding with the point at which the context window first includes the S-wave arrival and early surface-wave energy, and then saturates for $r \geq 1.5$. This saturation indicates that beyond a threshold amount of post-S context, additional observations provide diminishing returns. The sharp transition at $r = 1$ supports the view that the Configuration~B degradation is an information-availability problem: the model needs to observe the emerging surface-wave regime before it can extrapolate it.

\subsection{Representative Waveform Forecasts}
\label{sec:representative_forecasts}

\begin{figure}[t]
\centering
\includegraphics[width=\linewidth]{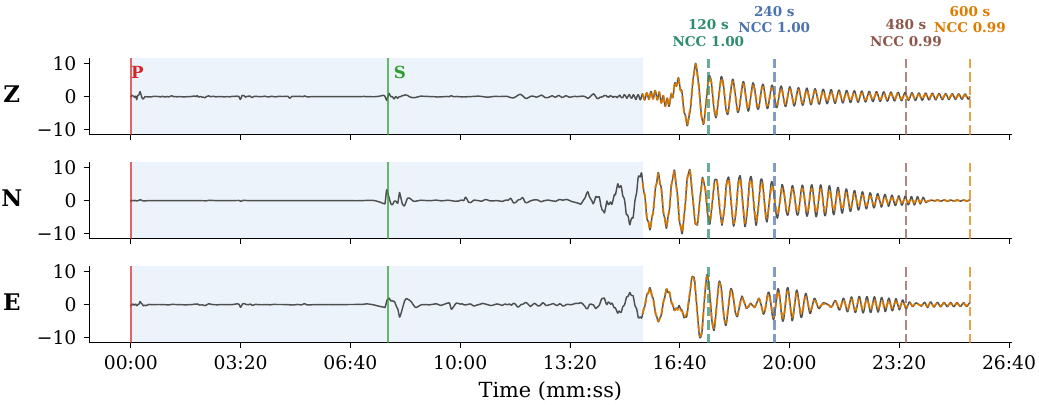}
\caption{Autoregressive forecast of a representative event ($M_w = 6.8$, $\Delta = 56^\circ$, depth = 55\,km) with context ratio $r = 2$. The ground-truth waveform is shown in dark gray and the model prediction in orange dashed. Vertical dashed lines mark four prediction horizons (120, 240, 480, 600\,s) with the global NCC evaluated up to each checkpoint. The model maintains NCC $\geq 0.99$ through the full 600\,s rollout across all three components.}
\label{fig:horizon_overlay}
\end{figure}

To provide physical intuition for the statistical results above, we examine a representative forecast in detail. Figure~\ref{fig:horizon_overlay} shows the autoregressive rollout for a single event ($M_w = 6.8$, $\Delta = 56^\circ$, depth = 55\,km) with the ground-truth waveform (dark) and the model prediction (orange dashed) overlaid on all three components. Vertical dashed lines mark four increasing prediction horizons at 120, 240, 480, and 600\,s, with the corresponding global NCC evaluated up to each checkpoint.

The model reproduces the observed waveform with high fidelity across the full 600\,s rollout, maintaining NCC values above 0.99 at all four checkpoints. During the first 120\,s of prediction, the forecast tracks the dominant body-wave coda and early surface-wave arrivals with near-perfect phase and amplitude agreement on all three components. As the rollout extends to 240 and 480\,s, the predicted waveform continues to follow the evolving surface-wave train, preserving both the envelope decay and the oscillatory structure of the coda. Even at 600\,s, which corresponds to 10 minutes of unsupervised autoregressive generation, the prediction remains phase-coherent with the ground truth, with no visible amplitude drift or spectral degradation.

This example is representative of the majority of test events: for the median event, the model produces stable, physically plausible continuations of the post-S wavefield over several hundred seconds. The key observation is that the prediction does not merely repeat a generic coda template but adapts to the specific radiation pattern, distance-dependent dispersion, and amplitude decay of the individual event, as evidenced by the close match across all three components simultaneously.

Not all events are predicted this well. The mean--median gap in NCC (Section~\ref{sec:overall_performance}) indicates a tail of events where the rollout degrades, typically through accumulated phase drift in the late coda. Representative failure cases are presented in Appendix~\ref{app:failure_modes}, where we show that the dominant failure mode is a gradual loss of phase coherence while the amplitude envelope and spectral content remain physically plausible. The model predicts a realistic-looking waveform that is simply out of phase with the true one.

\subsection{Limitations and Extensions}
\label{sec:limitations}

In this study, we have presented a controlled test case of how a causal transformer architecture can autoregressively forecast seismograms. Our test case is deliberately limited and the network as implemented in this study is several orders of magnitude smaller than modern large language models (LLMs) and contains only a fraction of the parameters typically associated with contemporary foundation models. Nevertheless, it already demonstrates stable waveform continuation and physically consistent spectral behavior. 

Recent advances in large-scale sequence modeling have shown that transformer performance follows predictable scaling laws with respect to model size, data volume, and compute \citep{Kaplan:2020trn}. The results reported here therefore likely represent a lower bound on achievable forecasting quality. We would expect substantial improvements in long-horizon stability, phase coherence, and spectral fidelity as model capacity and training scale increase in accordance with established scaling-law behavior. We therefore see promise in using an expanded \textsc{SeismoGPT} or similar transformer architectures for more complex seismic forecasting problems. As next steps, there are a multitude of ways in which \textsc{SeismoGPT} could be adapted and expanded. 

A logical next step would be to increase the range of parameters covered. The current dataset spans epicentral distances of 10--90$^\circ$ and source depths of 5--100\,km. Extending to deeper sources and distances beyond 90$^\circ$ would expose the model to fundamentally different wavefield character, where deep earthquakes excite fewer surface modes, producing more body-wave-dominated records, while the post-90$^\circ$ wavefield contains complex arrivals from transmission, reflection, and diffraction at the core--mantle and inner-core boundaries. 

A major expansion of \textsc{SeismoGPT} would be to incorporate real seismic data. Such a step poses considerable challenges. Firstly, the real Earth is substantially more complex than the 1D model used to calculate the synthetics in this implementation. Parts of Earth's interior, particularly the crust, lithosphere and lowermost mantle, are exceedingly heterogeneous, leading to variations in seismograms depending on the location of source and receiver. An intermediate step could be to adapt \textsc{SeismoGPT} to synthetics calculated using a 3D tomographic model of the mantle and crust  \citep[e.g.][]{french2014whole, lei2020global}. This poses practical problems, as it sacrifices the convenience of reciprocal Green's functions, and thus would require significant extra computational resources to generate the training data. It would also require the explicit definition of latitude and longitude when defining the sources and receivers.

Perhaps a greater challenge in moving to real data is the pervasive noise that permeates all seismic recordings, including ambient seismic noise, instrumental artifacts, and signals from other sources, which are entirely absent from our synthetic dataset. Additionally, real seismograms are typically sampled at much higher rates (e.g. 100\,Hz), so the current 16-sample token would correspond to only 160\,ms rather than the 8.4\,s it covers at the synthetic sampling rate, requiring experimentation with substantially larger token sizes. Addressing these challenges will likely require a combination of noise-robust training strategies and architectural adaptations.

Looking forward, an important limitation of the current approach is the use of fixed patch-based tokenization. Simple waveform patching effectively creates a near-infinite vocabulary in which each token corresponds to a numerical realization rather than a semantic representation of seismic dynamics. This limits the model’s ability to form higher-level abstractions and may contribute to phase drift during long-horizon rollout. Future work will therefore focus on more structured tokenization strategies, including learned discrete latent representations, multi-resolution tokens, and physics-informed embeddings that explicitly separate phase, envelope, and spectral characteristics. Such developments may enable semantic representations of propagation regimes and improve high-fidelity waveform forecasting over extended prediction horizons, particularly when combined with larger-scale models that leverage the scaling behavior observed in modern sequence learning systems.


\section{Conclusion}
We have presented SeismoGPT, a causal transformer for autoregressive forecasting of three-component seismograms. Trained on synthetic waveforms spanning source depths of 5--100\,km, epicentral distances of 10--90$^\circ$, and magnitudes $3 \leq M_w \leq 7$, the model achieves median NCC values between 0.93 and 0.97 across three evaluation configurations with prediction horizons of 120 and 240\,s. Performance depends on physically interpretable parameters, with forecast quality degrading at large distances and low magnitudes where wavefields are weakly coherent and highly dispersive. When the model fails, it produces physically plausible but phase-shifted waveforms rather than unphysical signals.

These results demonstrate that transformer-based sequence models can learn stable dynamical continuation of seismic wavefields from data alone, without explicit integration of the governing equations. We see this as a promising first step toward data-driven seismic forecasting, with potential applications in earthquake early warning, seismic hazard assessment, and ambient noise characterization for next-generation gravitational-wave observatories.

\section{Code and Data Availability}
The implementation of the \textsc{SeismoGPT} framework, including data generation, training, and evaluation pipelines, is publicly available as open-source software \footnote{https://github.com/wesmail/seismogpt}. The repository provides all components required to reproduce the experiments presented in this work, including synthetic waveform generation based on \texttt{Instaseis}, data handling utilities, model architectures implemented in \texttt{PyTorch Lightning} \cite{falcon2019pytorch}, and
quality-assurance scripts for autoregressive rollout evaluation and metric visualization. A persistent archived release is available via Zenodo at \url{https://doi.org/10.5281/zenodo.21044155}.

\section{Acknowledgment}
This work is funded by the ErUM-WAVE project 05D2022, "ErUM-Wave: Antizipation 3-dimensionaler Wellenfelder", which is supported by the German Federal Ministry of Research, Technology and Space (BMFTR). The work of WE is partially supported by the Science, Technology and Innovation Funding Authority (STDF) under grant number 50806. SR was funded by DFG grant TH1530/16-3.

\newpage

\appendix

\section{Representative Failure Cases}
\label{app:failure_modes}

\begin{figure}[!htb]
\centering
\includegraphics[width=\linewidth]{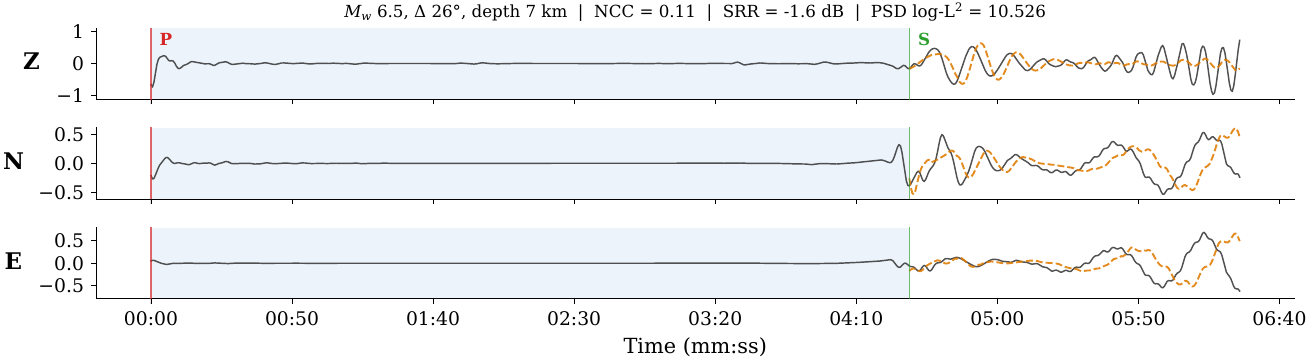}
\caption{Representative failure case for Configuration~A. Shallow intermediate-distance event (depth = 7.2\,km, $M_w = 6.52$, $\Delta = 25.6^\circ$). The predicted waveform remains oscillatory but loses phase coherence with the true waveform shortly after the context boundary, producing a global NCC of 0.11.}
\label{fig:fail_a}
\end{figure}

\begin{figure}[!htb]
\centering
\includegraphics[width=0.75\linewidth]{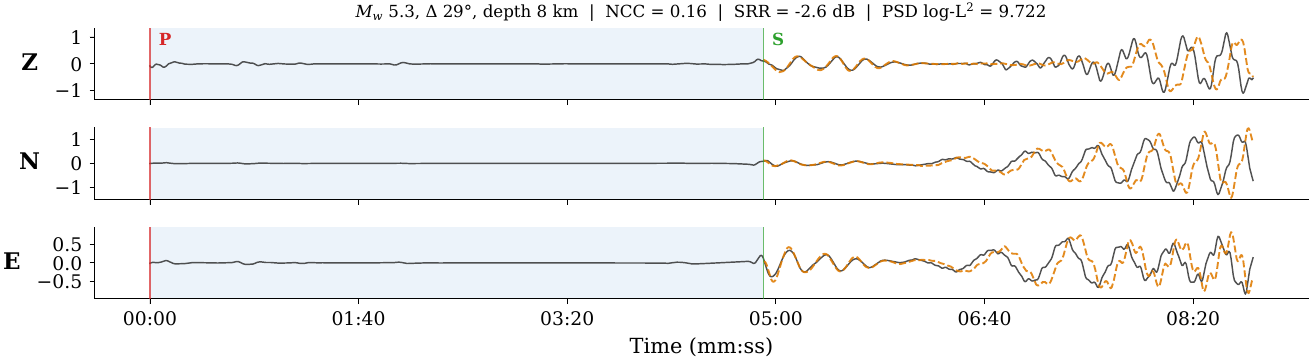}
\caption{Representative failure case for Configuration~B. Shallow intermediate-distance event (depth = 8.6\,km, $M_w = 5.30$, $\Delta = 29.1^\circ$). The early forecast is broadly aligned with the true waveform but accumulates phase and amplitude errors over the 240\,s rollout, producing a global NCC of 0.16.}
\label{fig:fail_b}
\end{figure}

\begin{figure}[!htb]
\centering
\includegraphics[width=0.75\linewidth]{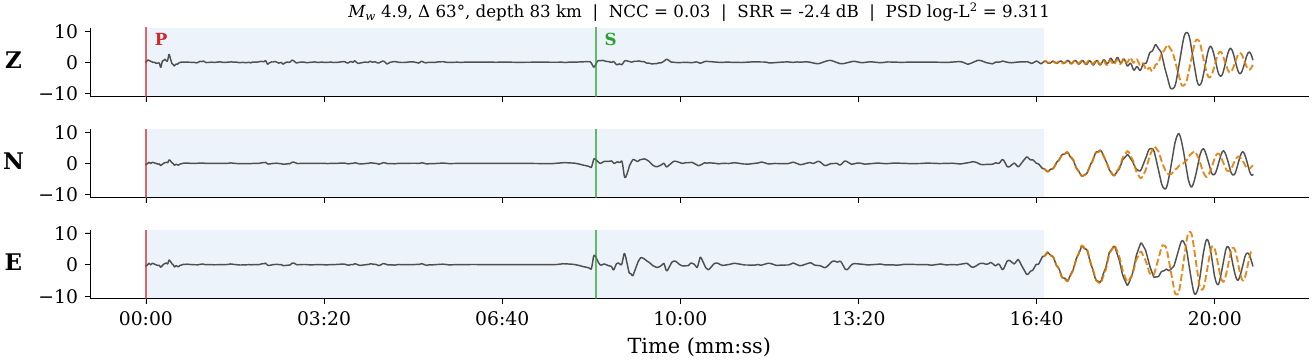}
\caption{Representative failure case for Configuration~C. Deep, large-distance event (depth = 82.6\,km, $M_w = 4.9$, $\Delta = 63.0^\circ$). Despite extended context, the forecast loses phase coherence with the true late-arriving energy, producing a global NCC of 0.03.}
\label{fig:fail_c}
\end{figure}

Figures~\ref{fig:fail_a}--\ref{fig:fail_c} show
representative failure cases for each configuration, selected from events with global NCC below the 5th percentile of the test-set distribution. In all three cases, the predicted waveform remains oscillatory and physically plausible, the model does not produce unphysically large amplitudes, divergent growth, or high-frequency noise. Instead, the dominant failure mode is phase mismatch: the prediction drifts out of alignment with the true waveform while maintaining a realistic envelope and spectral character. This distinguishes autoregressive forecasting failures from the numerical instabilities commonly seen in recurrent architectures, and suggests that the model has learned a valid dynamical regime but follows an incorrect trajectory within it.

Figure~\ref{fig:fail_a} shows a failure case for Configuration~A: a shallow event at intermediate distance (depth = 7.2\,km, $M_w = 6.52$, $\Delta = 25.6^\circ$). This is a relatively large earthquake, demonstrating that the failure is not simply a low-magnitude effect. The predicted continuation remains oscillatory and broadly consistent with the expected post-S wavefield, but it loses phase alignment with the true waveform almost immediately after the start of the autoregressive rollout. The resulting global NCC is 0.11 despite the absence of any obvious unphysical behavior. Notably, this event falls outside the primary failure corner identified in the parameter-plane analysis (Section~\ref{sec:results_geometry}), indicating that isolated phase-decorrelation failures can occur even in otherwise favorable parameter regimes, likely due to sensitivity to specific radiation-pattern or source-mechanism characteristics that are not captured by distance, depth, and magnitude alone.

Figure~\ref{fig:fail_b} shows a failure case for Configuration~B: a shallow, intermediate-distance event (depth = 8.6\,km, $M_w = 5.30$, $\Delta = 29.1^\circ$). Unlike the Configuration~A example, the forecast is not immediately unstable: the early post-S continuation remains oscillatory and broadly aligned with the true waveform. However, as the autoregressive rollout proceeds over the longer 240\,s prediction horizon, small timing and amplitude errors accumulate, leading to substantial late-time decorrelation across all three components. The resulting low NCC of 0.16 and SRR of $-2.6$, therefore reflect long-horizon phase and amplitude divergence rather than a failure to generate a physically plausible seismic waveform. This behavior is consistent with the dominant Configuration~B failure mode identified in the parameter-plane analysis: the model is released after only $1 \times (t_S - t_P)$ of post-S context but must sustain coherent prediction over a window long enough for small errors to compound.

Figure~\ref{fig:fail_c} shows a failure case for Configuration~C: a deep, large-distance event (depth = 82.6\,km, $M_w = 4.9$, $\Delta = 63.0^\circ$). Although Configuration~C provides an extended $2 \times (t_S - t_P)$ post-S context before the 240\,s rollout, the prediction loses phase coherence with the true late-arriving wave packet. The forecast remains oscillatory and physically plausible, but the detailed timing and amplitude evolution diverge from the target, producing a global NCC of 0.03 and SRR of $-2.4$. This case is particularly informative because the source magnitude is relatively large, so the failure cannot be attributed to a weak signal. The combination of large distance and deep source places this event in the shifted failure region identified in the Configuration~C parameter-plane analysis (Figure~\ref{fig:corner_c}), where the extended context window approaches the model's token budget and the predicted segment contains highly dispersive energy whose phase evolution is sensitive to small timing errors.

\section{Parameter-Plane Metrics for Configurations A and C}
\label{app:heatmaps}

\begin{figure*}[!bht]
\centering
\includegraphics[width=\textwidth]{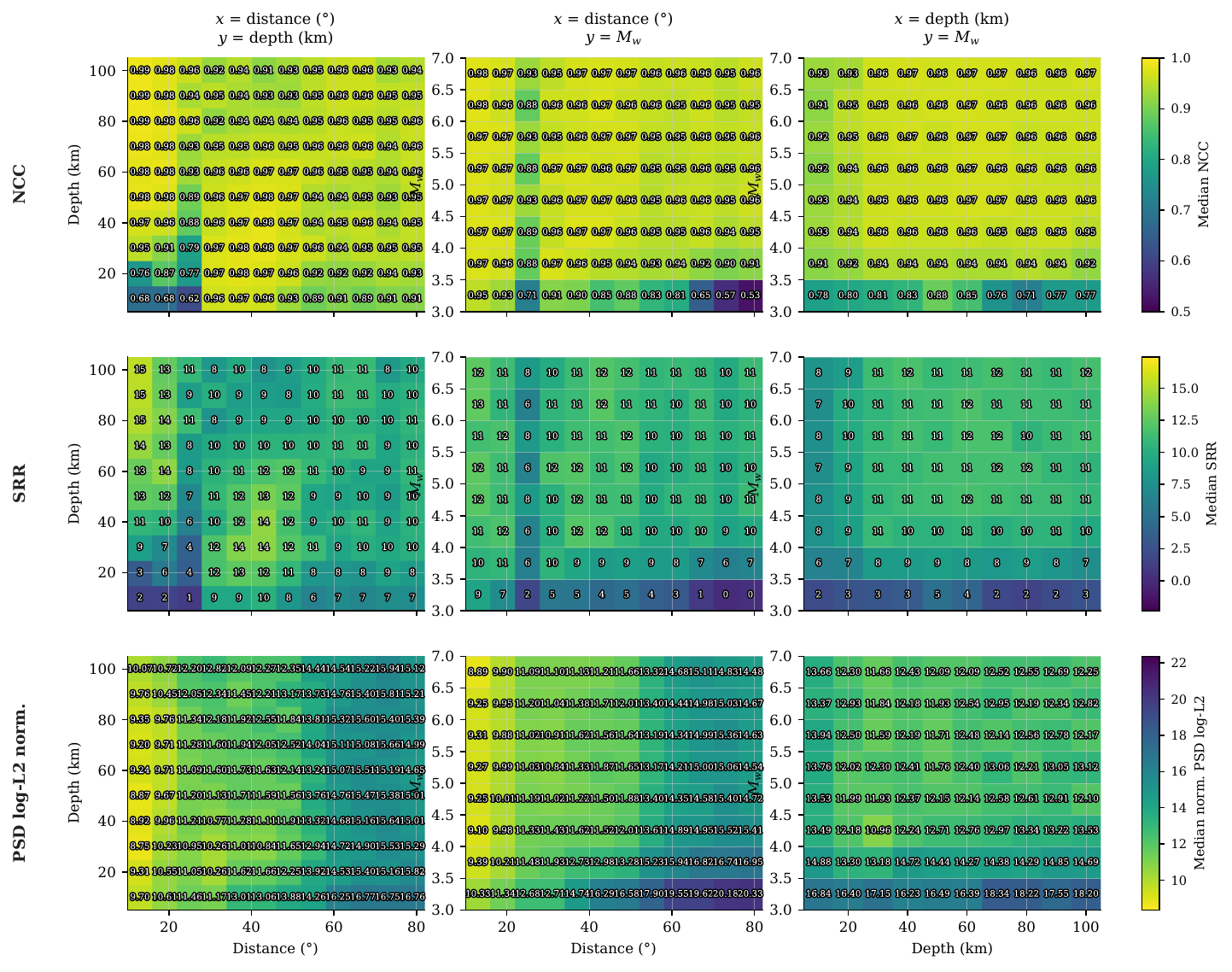}
\caption{Two-dimensional parameter-plane maps of median NCC (top), SRR (middle), and normalized PSD log-$L^2$ error (bottom) for Configuration~A.}
\label{fig:corner_a}
\end{figure*}

\begin{figure*}[!bht]
\centering
\includegraphics[width=\textwidth]{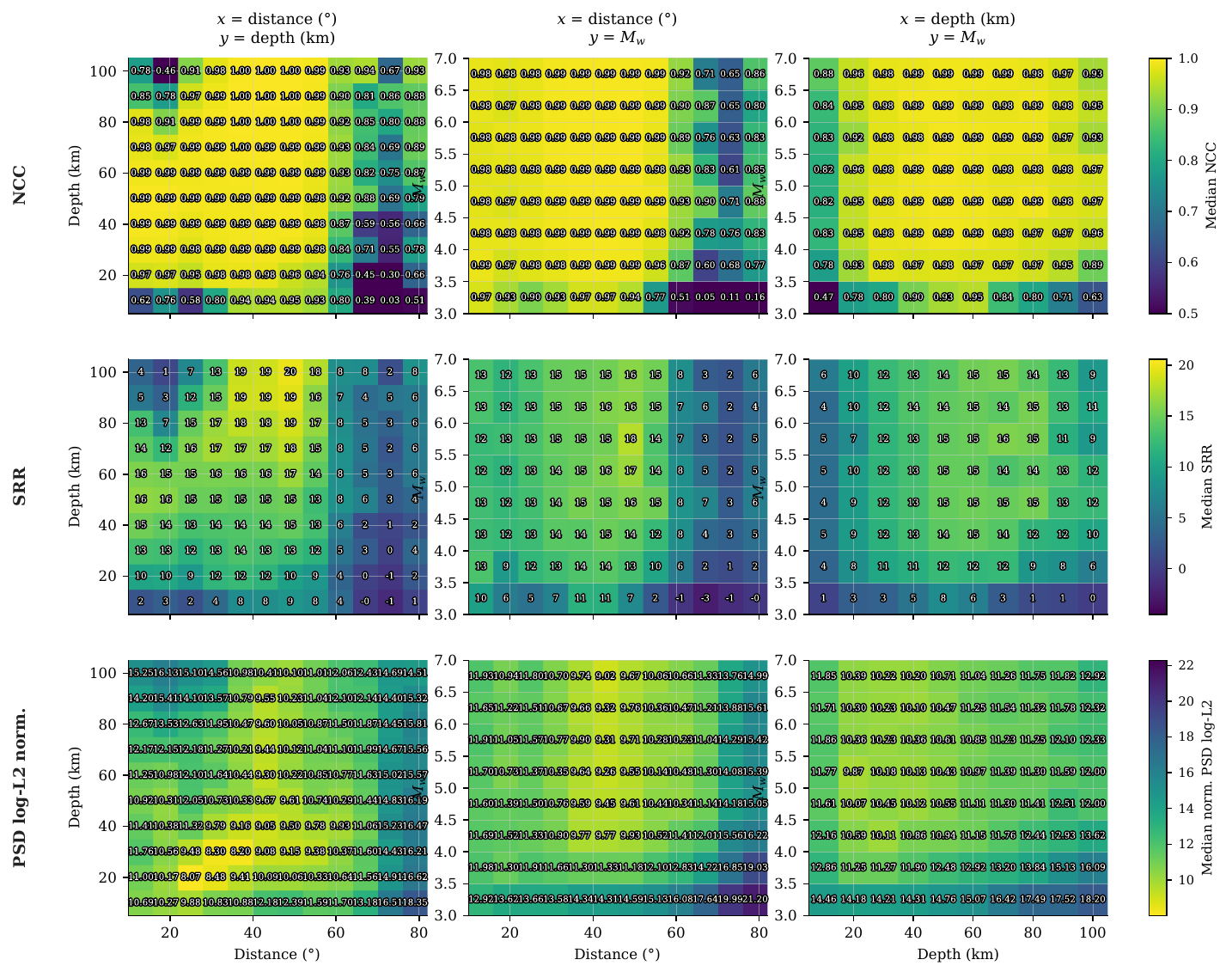}
\caption{Two-dimensional parameter-plane maps of median NCC (top), SRR (middle), and normalized PSD log-$L^2$ error (bottom) for Configuration~C.}
\label{fig:corner_c}
\end{figure*}

Figures~\ref{fig:corner_a} and \ref{fig:corner_c} show two-dimensional parameter-plane maps of median NCC, SRR, and PSD log-$L^2$ error for Configurations~A and C, complementing the Configuration~B analysis in Section~\ref{sec:results_geometry}.

Configuration~A (Figure~\ref{fig:corner_a}), with its shorter 120\,s prediction horizon, shows uniformly high NCC values ($\geq 0.90$) across most of the parameter space. The failure corner identified in Configuration~B, large distance, low magnitude, shallow depth, is still visible but considerably attenuated, with the lowest median NCC values remaining above 0.58. This confirms that the shorter prediction horizon limits the opportunity for autoregressive error accumulation, even in physically challenging regimes.

Configuration~C (Figure~\ref{fig:corner_c}), which doubles the context to $2 \times (t_S - t_P)$ while keeping the same 240\,s prediction horizon as B, recovers performance across most of the parameter space. However, a distinct failure region emerges at large distances ($\Delta \gtrsim 70^\circ$) combined with deep sources ($d \gtrsim 80$\,km), where median NCC drops sharply. This region differs from the low-magnitude/shallow corner that dominates Configuration~B, suggesting that at the largest distances and depths the extended context window approaches the model's token budget, potentially truncating the available context and limiting forecast quality.

\newpage\clearpage

\renewcommand\refname{References}
\bibliography{references.bib}
\newpage

\end{document}